\documentclass[aps,preprint]{revtex4}%
\usepackage{amsfonts}
\usepackage{amsmath}
\usepackage{amssymb}
\usepackage{graphicx}%
\providecommand{\U}[1]{\protect\rule{.1in}{.1in}}
\newcommand{\beq}{\begin{equation}}
\newcommand{\eeq}{\end{equation}}
\newcommand{\bea}{\begin{eqnarray}}
\newcommand{\eea}{\end{eqnarray}}
\begin{document}
\preprint{ }
\title{Revised Thomas-Fermi Approximation for Singular Potentials }
\author{James W.\ Dufty and S.B.\ Trickey}
\affiliation{Department of Physics, University of Florida, Gainesville FL 32611}

\begin{abstract}
Approximations to the many-fermion free energy density functional that include
the Thomas-Fermi (TF) form for the non-interacting part lead to singular
densities for singular external potentials (\textit{e.g.} attractive Coulomb).
This limitation of the TF approximation is addressed here by a formal map of
the exact Euler equation for the density onto an equivalent TF form
characterized by a modified Kohn-Sham potential. It is shown to be a
\textquotedblleft regularized\textquotedblright\ version of the Kohn-Sham
potential, tempered by convolution with a finite-temperature response
function. The resulting density is non-singular, with the equilibrium
properties obtained from the total free energy functional evaluated at this
density. This new representation is formally exact. Approximate expressions
for the regularized potential are given to leading order in a non-locality
parameter and the limiting behavior at high and low temperatures is described.
The non-interacting part of the free energy in this approximation is the usual
Thomas-Fermi functional. These results generalize and extend to finite
temperatures the ground-state regularization by Parr and Ghosh
(Proc.\ Nat.\ Acad.\ Sci.\ \textbf{83}, 3577 (1986)) and by Pratt, Hoffman,
and Harris (J.\ Chem.\ Phys. \textbf{92}, 1818 (1988)) and formally
systematize the finite-temperature regularization given by the latter authors.

\end{abstract}
\date{27 June 2016; revised 27 July 2016}
\maketitle

\section{Introduction}

\label{sec1}

Interest in orbital-free density functional theory (OFDFT) has revived
recently, driven by the unfavorable (polynomial in the number of states) cost
scaling of conventional Kohn-Sham (KS) calculations used to drive \emph{ab
initio} molecular dynamics simulations of ion-population thermodynamics. High
temperatures (compared to ambient), inherent to warm dense matter simulations,
worsen the cost problem. Details are irrelevant here; discussion with
extensive references can be found in Ref.\ \cite{IPAMReview}. Efforts to
improve approximate OFDFT functionals inexorably involve consideration of
known, well-characterized approximations. Of those, the Thomas-Fermi (TF)
model \cite{Thomas,Fermi} is, without doubt, the most thoroughly studied in
both physics and mathematics
\cite{LiebSimon77,Lieb81,March83,GoldsteinRieder89,Spruch91,BenilanBrezis}.

Improper limiting behavior of the electron number density $n_{TF}({\mathbf{r}%
})$ is among the defects of the TF scheme when applied to atoms. At $T=0$ K,
for an atom of nuclear charge $Z$, elementary TF (no exchange or correlation)
gives
\begin{equation}
n_{TF}({\mathbf{r}})=\left\{  {\large \left(  \frac{3}{5c_{TF}}\right)
\left[  \mu+\frac{Ze^{2}}{r}-v_{H}({\mathbf{r}})\right]  }\right\}  ^{3/2}\;,
\label{nTFatTeq0}%
\end{equation}
with $\mu$ the chemical potential, $v_{H}({\mathbf{r}})$ the Hartree
potential,
\begin{equation}
c_{TF}=\frac{3}{10m}\left(  \frac{3h^{3}}{8\pi}\right)  ^{2/3}\;,
\label{cTFdefin}%
\end{equation}
$m$ the electron mass, and $e$ the electron charge magnitude. It is seen that
$n_{TF}({\mathbf{r}})\sim r^{-3/2}$ as $\mathbf{r\rightarrow0}$. In contrast,
the correct behavior at the origin is non-singular
\cite{Kato57,Bingel63,PackByersBrown66,March..VanDoren2000,KryachkoLudena},
\begin{equation}
n({\mathbf{r}})\sim(1-2Zr/a_{0})+{\mathrm{O}}(r^{2})\;, \label{denscusp}%
\end{equation}
with $a_{0}=h^{2}/\pi me^{2}$ the Bohr radius. This difficulty extends to
current problems of interest for systems of electrons and positive ions
comprising warm dense matter.

Attention to this difficulty dates at least to the Scott correction
\cite{Scott52}, which Schwinger \cite{Schwinger80} rationalized by replacing
the near-nucleus TF density with a Bohr-atom-like density grafted smoothly to
$n_{TF}$ in the exterior region. More recent work falls into two groups,
schemes that modify the non-interacting energy or free energy with additive
corrections to the TF result, and those that modify the Euler equation for the
density with additive corrections to the
%\newline
TF Euler equation. In the former group are those approaches based on linear
combinations of TF and von Weizs\"acker \cite{Weizsacker} free energy
functionals, e.g.\ Ref.\ \cite{PlumertGeldart83}, or ad hoc constraints on the
density, e.g.\ Ref.\ \cite{PG86,GhoshParr87,MembradoPacheco90,GhoshDeb94}.
Phenomenological modifications of the Euler equation have been explored
extensively in Ref.\ \cite{ChaiWeeks} by imposition of known exact limits. It
is argued that the Euler equation is a more appropriate point at which to
introduce approximations, rather than directly for the non-interacting energy
or free energy functional. However, Euler equation methods determine only the
density and a subsequent reconstruction of the corresponding non-interacting
free energy is required for the thermodynamics.

Here we start with a formal solution to the exact Euler equation (including
exchange and correlation) and write it in a TF representation by introduction
of an effective potential. This is the approach of
Ref.\ \cite{PrattHoffmanHarris} introduced at zero temperature and extended to
finite temperatures in Ref.\ \cite{HoffmanHarrisPratt}. Exact expressions for
both the density and non-interacting free energy are given in TF
representations. The density is given by the usual TF form in terms of the
Kohn-Sham (KS) potential renormalized by the ideal gas non-linear response
function. The relationship to other recent work at finite temperatures is
discussed. The simplest approximation to these representations is shown to
remove the problem of singular densities for singular potentials. The work is
primarily an elaboration and completion of the ideas in
Ref.\ \cite{HoffmanHarrisPratt}.

\section{Thomas-Fermi and the problem for singular potentials}

\label{sec2} The essential element of finite-$T$ DFT for electrons is a free
energy functional $F(\beta\mid\cdot)$ that becomes the thermodynamic free
energy when evaluated at the equilibrium density $n(r)$
\cite{Mermin65,DuftyTrickeyMolPhys2016}. That density is determined from
solution to the Euler equation
\begin{equation}
\frac{\delta F(\beta\mid n)}{\delta n(\mathbf{r})}=\mu(\mathbf{r}),
\label{2.1}%
\end{equation}
with $\mu_{e}$ the chemical potential, $\mu(\mathbf{r})\equiv\mu
_{e}-v(\mathbf{r})$ the local chemical potential, and $v(\mathbf{r})$ the
given external single-particle potential. The additional constraints of
positivity, $n(\mathbf{r})\geq0$, and total number of electrons $\int
d\mathbf{r}n(\mathbf{r})=N\left(  \mu_{e}\right)  $, are left implicit for the
moment. The KS approach separates the free energy functional into its
non-interacting functional $F^{(0)}(\beta\mid\cdot)$ and the remainder
$F^{(1)}(\beta\mid\cdot)$. Here the specific form for $F^{(1)}(\beta\mid
\cdot)$ is not required (nor is the typical further separation of it into a
Hartree mean-field contribution and the rest, the exchange-correlation free
energy). Equation (\ref{2.1}) thus can be written equivalently as
\begin{equation}
\frac{\delta F^{(0)}(\beta\mid n)}{\delta n(\mathbf{r})}=\mu_{KS}%
(\mathbf{r,}\beta\mid n), \label{2.2}%
\end{equation}
with $\mu_{KS}(\mathbf{r,}\beta\mid n)$ identified as
\begin{equation}
\mu_{KS}(\mathbf{r,}\beta\mid n)=\mu(\mathbf{r})-\frac{\delta F^{(1)}%
(\beta\mid n)}{\delta n(\mathbf{r})} \label{2.3}%
\end{equation}
If the chemical potential is extracted from $\mu_{KS}(\mathbf{r,}\beta\mid
n)$, the remainder is known as the (negative) KS potential
\begin{equation}
\mu_{KS}(\mathbf{r,}\beta\mid n)=\mu_{e}-v_{KS}(\mathbf{r,}\beta\mid
n),\hspace{0.2in}v_{KS}(\mathbf{r,}\beta\mid n)\equiv v(\mathbf{r}%
)+\frac{\delta F^{(1)}(\beta\mid n)}{\delta n(\mathbf{r})}\;. \label{2.3a}%
\end{equation}

As recognized by Kohn and Sham \cite{KohnSham}, (\ref{2.2}) can be interpreted
as the thermodynamic derivative for a non-interacting system in the presence
of a local chemical potential $\mu_{KS}(\mathbf{r,}\beta\mid n)$. The
corresponding density is therefore (see Appendix \ref{apa} for the definitions
of the exact quantum representations for the non-interacting system)
\begin{align}
n(\mathbf{r})  &  =\left\langle \mathbf{r}\right\vert \left(  e^{\left(
\beta\left(  \frac{\widehat{p}^{2}}{2m}-\mu_{KS}(\mathbf{\widehat{\mathbf{q}%
},}\beta\mid n)\right)  \right)  }+1\right)  ^{-1}\left\vert \mathbf{r}%
\right\rangle \;\nonumber\\
&  \equiv n(\mathbf{r},\beta\mid\mu_{KS}), \label{2.4}%
\end{align}
where $\left\langle \mathbf{r}\right\vert \widehat{X}\left\vert \mathbf{r}%
\right\rangle $ denotes a single-particle matrix element in coordinate
representation. In general, a caret over a symbol denotes it as a
single-particle operator rather than one of its eigenvalues. The notation in
(\ref{2.4}) makes the dependence of the density upon the external potential
explicit, an explicitness that will be useful in what follows. Evidently the
condition $n(\mathbf{r})\geq0$ is satisfied, and the condition $\int
d\mathbf{r}n(\mathbf{r})=N\left(  \mu_{e}\right)  $ is enforceable by the
choice for $\mu_{e}$ whenever solutions to (\ref{2.4}) exist. Equation
(\ref{2.4}) is entirely equivalent to the Euler equation (\ref{2.2}) for the
determination of the density.

Of course, the exact ``solution'' (\ref{2.4}) is only implicit since $\mu
_{KS}(\mathbf{r,}\beta\mid n)$ is a functional of the density through its
dependence on $F^{(1)}$. Furthermore, since $\mu_{KS}(\mathbf{\widehat
{\mathbf{q}},}\beta\mid n)$ appears as a function of the coordinate operator
$\widehat{\mathbf{q}}$, the right side of (\ref{2.4}) is a non-local
functional of the density, beyond the dependence intrinsic to $F^{(1)}%
(\beta\mid n)$ (because of the non-commuting operators $\widehat{\mathbf{p}}$
and $\widehat{\mathbf{q}}$). A local approximation, wherein $\mu
_{KS}(\mathbf{\widehat{\mathbf{q}},}\beta\mid n)$ is replaced by $\mu
_{KS}(\mathbf{r,}\beta\mid n)$ with $\mathbf{r}$ the same point as occurs on
the left side of (\ref{2.4}), leads to the familiar Thomas-Fermi approximation
(now with full exchange and correlation).
\begin{equation}
n_{TF}(\beta,\mu_{KS}(\mathbf{r}))=\frac{2}{h^{3}}\int d\mathbf{p}\left(
e^{\left(  \beta\left(  \frac{p^{2}}{2m}-\mu_{KS}(\mathbf{r,}\beta\mid
n)\right)  \right)  }+1\right)  ^{-1}\;.\label{2.5}%
\end{equation}
From this expression it is clear that $\mu_{KS}(\mathbf{r,}\beta\mid n)$ has
the same relationship to the density as for the uniform ideal Fermi gas
chemical potential $\mu^{(0)}(\beta,n)$. Thus, an equivalent form for the
Euler equation in the conventional TF approximation is
\begin{equation}
\mu^{(0)}(\beta,n_{TF}\left(  \mathbf{r}\right)  ) \approx \mu_{KS}%
(\mathbf{r,}\beta\mid n_{TF}).\label{2.8}%
\end{equation}
The corresponding TF non-interacting functional $F_{TF}^{(0)}(\beta\mid n)$ is
(see Appendix \ref{apa})
\begin{equation}
F_{TF}^{(0)}(\beta\mid n)=\int d\mathbf{r}f_{TF}^{(0)}(\beta,n\left(
\mathbf{r}\right)  ),\label{2.6}%
\end{equation}%
\begin{equation}
f_{TF}^{(0)}(\beta,n\left(  \mathbf{r}\right)  )\equiv-\frac{2}{\beta h^{3}%
}\int d\mathbf{p}\ln\left(  1+e^{-\beta(\frac{p^{2}}{2m}-\mu^{(0)}%
(\beta,n\left(  \mathbf{r}\right)  )}\right)  +\mu^{(0)}(\beta,n\left(
\mathbf{r}\right)  )n(\mathbf{r}).\label{2.7}%
\end{equation}
The approximate Euler equation obtained from (\ref{2.2}) by the replacement
$F^{(0)}(\beta\mid n)\rightarrow F_{TF}^{(0)}(\beta\mid n)$ is
\begin{equation}
\frac{\delta F_{TF}^{(0)}(\beta\mid n)}{\delta n(\mathbf{r})}=\mu
_{KS}(\mathbf{r,}\beta\mid n).\label{2.9}%
\end{equation}
It is easily checked that evaluation of the left side in (\ref{2.9}) gives
back (\ref{2.8}) (as an equality). 
The explicit equations (\ref{2.5}) or (\ref{2.8}) for the
density, and the corresponding expression for $F_{TF}^{(0)}$ are the simplest
example of an ``orbital free'' DFT (assuming that $F^{(1)}(\beta\mid n)$ is also
given in orbital free form).

Consider an attractive external potential which is singular at $\mathbf{r}=0$
\textit{e.g.}, the electron-ion Coulomb interaction. Then $\mu_{KS}%
(\mathbf{r,}\beta\mid n)\rightarrow\infty$ as $\mathbf{r\rightarrow0}$. In
turn, that implies $n\left(  \mathbf{r}\right)  \rightarrow\infty$ as
$\mathbf{r\rightarrow0}$ (further details are provided in Appendix \ref{apb}).
This unphysical consequence of the local approximation is precisely the
singularity discussed in the Introduction, recovered here in the $T\geq0$ K
context. The representations of the next sections provide a natural removal of
that singularity.

\section{Formally exact Thomas-Fermi-type representations.}

\label{sec3} In this section Thomas-Fermi-type representations are defined for
the solution to the Euler equation (\ref{2.4}) giving the equilibrium density
$n\left(  \mathbf{r}\right)  $, and for the non-interacting free energy
$F^{(0)}(\beta\mid n)$ evaluated at that density. The results are exact but
formal. Simple practical approximations are provided in the subsequent sections.

\subsection{Representation for $n\left(  \mathbf{r}\right)  $}

A motivation for the following discussion is provided by extracting the TF
free energy explicitly from $F^{(0)}(\beta\mid n)$%
\begin{equation}
F^{(0)}=F_{TF}^{(0)}+\Delta F^{(0)}.\label{3.1}%
\end{equation}
Since the TF free energy results from a ``local density approximation'', $\Delta
F^{(0)}$ represents the non-local contribution to $F^{(0)}(\beta\mid n)$.
Then, we can write  the exact Euler equation (\ref{2.2}) in the equivalent
form%
\begin{equation}
\frac{\delta F_{TF}^{(0)}(\beta\mid n)}{\delta n(\mathbf{r})}=\mu
_{KS}(\mathbf{r,}\beta\mid n)-\frac{\delta\Delta F^{(0)}(\beta\mid n)}{\delta
n(\mathbf{r})}.\label{3.2}%
\end{equation}
Note the distinction with the conventional TF approximation, 
equation (\ref{2.9}). One thus
sees that the exact solution to the Euler equation can be given the TF form,
but with an effective external potential%
\begin{equation}
n(\mathbf{r})=n_{TF}(\beta,z(\mathbf{r}))=\frac{2}{h^{3}}\int d\mathbf{p}%
\left(  e^{\left(  \beta\left(  \frac{p^{2}}{2m}-z(\mathbf{r})\right)
\right)  }+1\right)  ^{-1}\;.\label{3.3}%
\end{equation}
The effective potential $z(\mathbf{r})$ is identified as
\begin{equation}
z(\mathbf{r,}\beta\mid\mu_{KS})=\mu_{KS}(\mathbf{r,}\beta\mid n)-\frac
{\delta\Delta F^{(0)}(\beta\mid n)}{\delta n(\mathbf{r})}.\label{3.4}%
\end{equation}
A more constructive identification is obtained from the equality of
(\ref{2.4}) and (\ref{3.3})%
\begin{equation}
\left\langle \mathbf{r}\right\vert \left(  e^{\left(  \beta\left(
\frac{\widehat{p}^{2}}{2m}-\mu_{KS}(\mathbf{\widehat{\mathbf{q}},}\beta\mid
n)\right)  \right)  }+1\right)  ^{-1}\left\vert \mathbf{r}\right\rangle
\equiv\frac{2}{h^{3}}\int d\mathbf{p}\left(  e^{\left(  \beta\left(
\frac{p^{2}}{2m}-z(\mathbf{r})\right)  \right)  }+1\right)  ^{-1}.\label{3.4a}%
\end{equation}

Since (\ref{3.3}) has the same form as for the uniform ideal Fermi gas,
$z(\mathbf{r})$ has the same relationship to the density
\begin{equation}
\mu^{(0)}(\beta,n\left(  \mathbf{r}\right)  )=z(\mathbf{r}), \label{3.2a}%
\end{equation}
similar to the TF result (\ref{2.8}) except with $\mu_{KS}(\mathbf{r,}%
\beta\mid n)$ replaced by $z(\mathbf{r})$. More explicitly, this relationship
is determined from the dimensionless form
\begin{equation}
\lambda^{3}n(\mathbf{r})=2f_{3/2}(e^{\beta\mu^{(0)}(\beta,n\left(
\mathbf{r}\right)  )})\;, \label{3.2b}%
\end{equation}
where $\lambda=\left(  2\pi\beta\hbar^{2}/m\right)  ^{1/2}$ is the thermal de
Broglie wavelength and $f_{3/2}(x)$ is the Fermi integral \cite{HuangText}
\begin{equation}
f_{3/2}(x)=\frac{4}{\sqrt{\pi}}\int_{0}^{\infty}dyy^{2}\left(  x^{-1}e^{y^{2}%
}+1\right)  ^{-1}. \label{3.2c}%
\end{equation}
Accurate analytic fits to $f_{3/2}(x)$ and its inverse are available
\cite{Blakemore1982}.

More significantly, (\ref{3.4a}) defines $z(\mathbf{r})\equiv z(\mathbf{r,}%
\beta\mid\mu_{KS})$ as a \emph{functional} of $\mu_{KS}(\mathbf{r})$.
Construction of such a $z({\mathbf{r}},\beta\mid\mu_{KS})$, hence
demonstration of its existence, is the heart of the present problem. If
$z(\mathbf{r,}\beta\mid\mu_{KS})$ can be determined to arbitrarily accurate
approximation, its equality with $\mu^{(0)}(\beta,n\left(  \mathbf{r}\right)
)$ then provides an explicit equation to determine $n(r)$ for a given
$\mu_{KS}(\mathbf{r})$,
\begin{equation}
\mu^{(0)}(\beta,n\left(  \mathbf{r}\right)  )=z(\mathbf{r,}\beta\mid\mu
_{KS})\;, \label{3.5}%
\end{equation}
or, equivalently,
\begin{equation}
n(\mathbf{r})=\frac{2}{h^{3}}\int d\mathbf{p}\left(  e^{\left(  \beta\left(
\frac{p^{2}}{2m}-z(\mathbf{r,}\beta\mid\mu_{KS})\right)  \right)  }+1\right)
^{-1}\;. \label{3.6}%
\end{equation}
This is similar to the TF result (\ref{2.4}), but now (\ref{3.6}) is a
formally exact TF-type representation of the Euler equation through the
definition of $z(\mathbf{r,}\beta\mid\mu_{KS}).$

To illustrate a simple approximation to $z(\mathbf{r,}\beta\mid\mu_{KS})$ and
its consequences, return to (\ref{3.4}) and retain the leading contributions
to non-uniformity in $\Delta F^{(0)}$ \cite{Perrot79}%
\begin{equation}
z(\mathbf{r,}\beta\mid\mu_{KS})\rightarrow\mu_{KS}(\mathbf{r,}\beta\mid
n)+\frac{\hbar^{2}}{8m_{e}}\left[  \frac{\left\vert \nabla n(\mathbf{r}%
)\right\vert ^{2}}{n^{2}(\mathbf{r})}-2\frac{\nabla^{2}n(\mathbf{r}%
)}{n(\mathbf{r})}\right]  . \label{3.6e}%
\end{equation}
It can be shown that the term proportional to $\nabla^{2}n(\mathbf{r})$ leads
to non-singular solutions with this $z(\mathbf{r,}\beta\mid\mu_{KS})$ in
(\ref{3.5}) or (\ref{3.6}) for the case of an external positive ion. Hence
even the leading corrections to the local density approximation can cure the
problem with singular potentials.

Of course, determination of $z(\mathbf{r,}\beta\mid\mu_{KS})$ more generally
is as difficult as the original problem of (\ref{2.4}), namely, to determine
$n(\mathbf{r})\equiv n(\mathbf{r,}\beta\mid\mu_{KS})$. However, the
expectation here is that simple systematic approximations to $z(\mathbf{r,}%
\beta\mid\mu_{KS})$ directly from (\ref{3.1}) will give significantly better
approximations for $n(\mathbf{r,}\beta\mid\mu_{KS})$ than from conventional TF
theory (neglect of $\Delta F^{(0)}$ or phenomenological approximations to it).
For this purpose an appropriate exact representation for $z(\mathbf{r,}%
\beta\mid\mu_{KS})$ in terms of the non-linear response function for the ideal
Fermi gas is obtained in Appendix \ref{apc} with the result%
\begin{equation}
z(\mathbf{r,}\beta\mid\mu_{KS})=\int d\mathbf{r}^{\prime}G(\mathbf{r}%
,\mathbf{r}^{\prime}\mathbf{,}\beta\mid z,\Delta\mu_{KS})\mu_{KS}%
(\mathbf{r}^{\prime})\;. \label{3.10}%
\end{equation}
Here $G(\mathbf{r},\mathbf{r}^{\prime}\mathbf{,}\beta\mid z,\Delta\mu_{KS})$
is constructed from the ideal Fermi gas response functional for local chemical
potential $x(\mathbf{r})$
\begin{equation}
R(\mathbf{r},\mathbf{r}^{\prime}\mathbf{,}\beta\mid x)=\frac{\delta
n(\mathbf{r},\beta\mid x)}{\delta x\left(  \mathbf{r}^{\prime}\right)  }\;,
\label{3.11}%
\end{equation}
such that $G(\mathbf{r},\mathbf{r}^{\prime}\mathbf{,}\beta\mid z,\mu_{KS})$ is
the integral of this response function along a linear path in function space
from $z(\mathbf{r})$ to $\mu_{KS}(\mathbf{r^{\prime}})$
\begin{equation}
G(\mathbf{r},\mathbf{r}^{\prime}\mathbf{,\beta}\mid z,\mu_{KS})=\frac{\int
_{0}^{1}dxR(\mathbf{r},\mathbf{r}^{\prime}\mathbf{,}\beta\mid z+x\Delta
\mu_{KS})}{\int d\mathbf{r}^{\prime}\int_{0}^{1}dxR(\mathbf{r},\mathbf{r}%
^{\prime}\mathbf{,}\beta\mid z+x\Delta\mu_{KS})}\;. \label{3.12}%
\end{equation}
Clearly, $G$ is normalized to unity in the sense
\begin{equation}
\int d\mathbf{r}^{\prime}G(\mathbf{r},\mathbf{r}^{\prime}\mathbf{,}\beta\mid
z,\Delta\mu_{KS})=1\;. \label{3.13}%
\end{equation}
Finally, $\Delta\mu_{KS}(\mathbf{r}^{\prime}\mathbf{)}$ is the difference
between $\mu_{KS}\left(  \mathbf{r}^{\prime}\right)  $ at all points
$\mathbf{r}^{\prime}$ and $z\left(  \mathbf{r}\right)  $ at the chosen point
$\mathbf{r}$
\begin{equation}
\Delta\mu_{KS}(\mathbf{r}^{\prime}\mathbf{)}\equiv\mu_{KS}\left(
\mathbf{r}^{\prime}\right)  -z\left(  \mathbf{r}\right)  ,\text{ \ \ \ \ for
all }\mathbf{r}^{\prime} \label{3.14}%
\end{equation}

\subsection{Representation for $F^{(0)}(\beta\mid n)$}

Once the equilibrium density has been found it can be used in the given
functional $F^{(1)}(\beta\mid n)$ to get the interacting part of the
thermodynamic free energy. However, the non-interacting contribution
$F^{(0)}(\beta\mid n)$ remains unknown to this point. One possibility would be
to construct it exactly from (see Appendix \ref{apb})
\begin{equation}
F^{(0)}(\beta\mid n)=\int d\mathbf{r}\left[  -\beta^{-1}\left\langle
\mathbf{r}\right\vert \ln\left(  1+e^{-\beta\left(  \frac{\widehat{p}^{2}}%
{2m}-\mu_{KS}(\widehat{\mathbf{r}}\mathbf{,}\beta\mid n)(\widehat{\mathbf{r}%
})\right)  }\right)  \left\vert \mathbf{r}\right\rangle +\mu_{KS}%
(\mathbf{r,}\beta\mid n)n(\mathbf{r})\right]  , \label{3.14a}%
\end{equation}
in terms of the eigenfunctions and eigenvalues of the KS Hamiltonian
$p^{2}/2m+$ $v_{KS}(\mathbf{r,}\beta\mid n)$. This is a straightforward
calculation once $n$ has been determined.

However, in the context of an approximate evaluation of $z(\mathbf{r,}%
\beta\mid\mu_{KS})$, it is not clear that the approximate density from the
corresponding form of (\ref{3.3}) will be the same as that constructed from
the eigenvalues and eigenfunctions of the KS Hamiltonian. An alternative
approach is to construct the grand potential, $\Omega^{(0)}(\beta\mid\mu
_{KS})$, from its functional derivative (see Appendix \ref{apc})
\begin{equation}
\frac{\delta\Omega^{(0)}(\beta\mid\mu_{KS})}{\delta\mu_{KS}\left(
\mathbf{r}\right)  }=-n(\mathbf{r})=-n_{TF}(\beta,z(\mathbf{r,}\beta\mid
\mu_{KS})), \label{3.15}%
\end{equation}
and then to determine $F^{(0)}(\beta\mid n)$ from the Legendre transform
\begin{align}
F^{(0)}(\beta &  \mid n)\equiv\Omega^{(0)}(\beta\mid\mu_{KS})-\int
d\mathbf{r}\mu_{KS}(\mathbf{r})\frac{\delta\Omega^{(0)}(\beta\mid\mu_{KS}%
)}{\delta\mu_{KS}\left(  \mathbf{r}\right)  }\label{3.16a}\\
&  =\Omega^{(0)}(\beta\mid\mu_{KS})+\int d\mathbf{r}\mu_{KS}(\mathbf{r}%
)n_{TF}(\beta,z(\mathbf{r,}\beta\mid\mu_{KS})). \label{3.16b}%
\end{align}
The last equalities of (\ref{3.15}) and (\ref{3.16b}) have made use of the
exact TF representation for the density (\ref{3.3}). In this way, a TF-like
representation for $F^{(0)}(\beta\mid n)$ also is obtained in Appendix
\ref{apc}, with the results
%TCIMACRO{\TEXTsymbol{\backslash}}%
%BeginExpansion
%$\backslash$%
%EndExpansion%
\begin{equation}
F^{(0)}(\beta\mid n)=F_{TF}(\beta\mid n)+\int d\mathbf{r}d\mathbf{r}^{\prime
}d\mathbf{r}^{\prime\prime}h(\mathbf{r};\mathbf{r}^{\prime},\mathbf{r}%
^{\prime\prime},\beta\mid z,\Delta\mu_{KS})\Delta\mu_{KS}\left(
\mathbf{r}^{\prime}\right)  \Delta\mu_{KS}\left(  \mathbf{r}^{\prime\prime
}\right)  , \label{3.17}%
\end{equation}
where $h(\mathbf{r,r}^{\prime},\mathbf{r}^{\prime\prime},\beta\mid z,\Delta
\mu_{KS})$ is a second-order response function given by eq.\ (\ref{c.11}) of
Appendix \ref{apc}. Also, $F_{TF}(\beta\mid n)$ is the usual TF free energy,
(\ref{2.6}), now evaluated at the exact density $n$. The definition of
$\Delta\mu_{KS}$ is the same as in (\ref{3.14}).

Equations (\ref{3.6}) and (\ref{3.17}) with (\ref{3.10}) are the primary exact
reformulations of DFT thermodynamics in a language most similar to the
approximate TF results. The essential required input is the functional
$z(\mathbf{r,}\beta\mid\mu_{KS})$, representing non-local corrections to
$\mu_{KS}(\mathbf{r,}\beta\mid n)$ in the usual TF approximation. The next
section shows that the simplest such correction removes the singular density
problem of TF.

\section{Approximation}

\label{sec4}

The foregoing formal analysis is expressed in terms of $\Delta\mu_{KS}\left(
\mathbf{r}^{\prime}\right)  $. It can be written as%
\begin{equation}
\Delta\mu_{KS}\left(  \mathbf{r}^{\prime}\right)  =\mu_{KS}\left(
\mathbf{r}^{\prime}\right)  -z\left(  \mathbf{r}\right)  =-\int d\mathbf{r}%
^{\prime\prime}G(\mathbf{r},\mathbf{r}^{\prime\prime}\mathbf{,}\beta,n\left(
\mathbf{r}\right)  \mid\Delta\mu_{KS})\left(  \mu_{KS}(\mathbf{r}%
^{\prime\prime})-\mu_{KS}(\mathbf{r}^{\prime})\right)  , \label{7.1}%
\end{equation}
which is seen to vanish for uniform systems. More generally, in this section
it is considered to be small so as to obtain leading order approximations for
$z(\mathbf{r,}\beta\mid\mu_{KS})$ and $F^{(0)}(\beta\mid n)$,%
\begin{equation}
z(\mathbf{r,}\beta\mid\mu_{KS})\rightarrow\int d\mathbf{r}^{\prime
}G(\mathbf{r},\mathbf{r}^{\prime}\mathbf{,}\beta\mid z,\Delta\mu_{KS}%
=0)\mu_{KS}(\mathbf{r}^{\prime})\;, \label{7.2}%
\end{equation}%
\begin{equation}
F^{(0)}(\beta\mid n)\rightarrow F_{TF}(\beta\mid n). \label{7.3}%
\end{equation}
Note that this is not the same as a gradient expansion since all higher order
derivatives are retained in (\ref{7.2}). Also, the density in (\ref{7.2})\ is
determined from (\ref{3.3}) using (\ref{7.2}) so $F_{TF}(\beta\mid n)$ differs
from $F_{TF}(\beta\mid n_{TF})$ by all higher order derivatives as well.

At this leading order approximation%
\begin{equation}
G(\mathbf{r},\mathbf{r}^{\prime}\mathbf{,}\beta\mid z,\Delta\mu_{KS}%
=0)=G(\mid{\mathbf{r}}-\mathbf{r}^{\prime}\mid,\beta,n\left(  \mathbf{r}%
\right)  )=\frac{R(\left\vert \mathbf{r}-\mathbf{r}^{\prime}\right\vert
,\beta,z\left(  \mathbf{r}\right)  )}{\int d\mathbf{r}^{\prime\prime
}R(r^{\prime\prime},\beta,z\left(  \mathbf{r}\right)  )} \label{7.4}%
\end{equation}
where $R(\left\vert \mathbf{r}-\mathbf{r}^{\prime}\right\vert ,\beta,z\left(
\mathbf{r}\right)  )$ is the response function for the \emph{homogeneous }
ideal Fermi gas as a function of $z\left(  \mathbf{r}\right)  $,
\begin{equation}
R(\left\vert \mathbf{r}-\mathbf{r}^{\prime}\right\vert ,\beta,z\left(
\mathbf{r}\right)  )=\frac{\delta n(\mathbf{r},\beta\mid x)}{\delta x\left(
\mathbf{r}^{\prime}\right)  }\Big\vert_{x=z(\mathbf{r})}\;. \label{7.5}%
\end{equation}
Furthermore, from (\ref{3.2a}), $z\left(  \mathbf{r}\right)  $ can be replaced
by the known function of the density $\mu^{(0)}(\beta,n\left(  \mathbf{r}%
\right)  ),$
\begin{equation}
G(\mid{\mathbf{r}}-\mathbf{r}^{\prime}\mid,\beta,n\left(  \mathbf{r}\right)
)=\frac{R(\left\vert \mathbf{r-r}^{\prime}\right\vert ,\beta,\mu^{(0)}%
(\beta,n\left(  \mathbf{r}\right)  ))}{\int d\mathbf{r}^{\prime\prime
}R(r^{\prime\prime}\mathbf{,}\beta,\mu^{(0)}(\beta,n\left(  \mathbf{r}\right)
))}\;. \label{7.6}%
\end{equation}

This local linear response approximation for $z\left(  \mathbf{r}\right)  $
gives the desired practical forms%
\begin{equation}
z\left(  \mathbf{r},\beta\mid\mu_{KS}\right)  \rightarrow\int d\mathbf{r}%
^{\prime}G(\left\vert \mathbf{r}-\mathbf{r}^{\prime}\right\vert \mathbf{,}%
\beta,n\left(  \mathbf{r}\right)  )\mu_{KS}(\mathbf{r}^{\prime}\mathbf{,}%
\beta\mid n). \label{7.7}%
\end{equation}
The corresponding equilibrium density is the solution to a TF form, except
with a renormalized KS potential
\begin{equation}
n(\mathbf{r})=\frac{2}{h^{3}}\int d\mathbf{p}\left(  e^{\left(  \beta\left(
\frac{p^{2}}{2m}+\widetilde{v}_{KS}(\mathbf{r,}\beta\mid n)-\mu_{e}\right)
\right)  }+1\right)  ^{-1}\;, \label{7.8}%
\end{equation}%
\begin{equation}
\widetilde{v}_{KS}(\mathbf{r,}\beta\mid n)=\int d\mathbf{r}^{\prime
}G(\left\vert \mathbf{r}-\mathbf{r}^{\prime}\right\vert \mathbf{,}%
\beta,n\left(  \mathbf{r}\right)  )v_{KS}(\mathbf{r}^{\prime}\mathbf{,}%
\beta\mid n). \label{7.9}%
\end{equation}
The non-interacting TF free energy still is given by (\ref{2.6}) and
(\ref{2.7})
\begin{equation}
F_{TF}^{(0)}(\beta\mid n)=\int d\mathbf{r}\left(  -\frac{2}{\beta h^{3}}\int
d\mathbf{p}\ln\left(  1+e^{-\beta(\frac{p^{2}}{2m}-\mu^{(0)}(\beta,n\left(
\mathbf{r}\right)  )}\right)  +\mu^{(0)}(\beta,n\left(  \mathbf{r}\right)
)n(\mathbf{r})\right)  , \label{7.10}%
\end{equation}
with now the density determined from (\ref{7.8}) instead of (\ref{2.5}).
Equations (\ref{7.8}) - (\ref{7.10}) comprise the simplest practical
application of the present analysis.

The function $G(r\mathbf{,}\beta,n\left(  \mathbf{r}\right)  )$ is calculated
in Appendix \ref{apc}, with the result
\begin{equation}
G(r\mathbf{,}\beta,n)=\frac{1}{\lambda\sqrt{\pi}}\frac{1}{r^{2}}\int
_{0}^{\infty}dx\,xp(x,\beta,\mu^{(0)}(\beta,n\left(  \mathbf{r}\right)
))\sin\left(  x\frac{4\sqrt{\pi}r}{\lambda}\right)  \;, \label{3.20}%
\end{equation}%
\begin{equation}
p(x,\beta,\mu^{(0)})\equiv\frac{\left(  e^{-\beta\mu^{(0)}}e^{x^{2}}+1\right)
^{-1}}{\int_{0}^{\infty}dx\left(  e^{-\beta\mu^{(0)}}e^{x^{2}}+1\right)
^{-1}}\;. \label{3.21}%
\end{equation}
Note that all of the density dependence of $G(r\mathbf{,}\beta,n)$ occurs
through $\mu^{(0)}(\beta,n\left(  \mathbf{r}\right)  ).$ The non-degenerate
limit of $G(\mid\mathbf{r}-\mathbf{r}^{\prime}\mid,\beta,n\left(
\mathbf{r}\right)  )$ occurs at high temperatures or low densities for which
$\lambda^{3}n<<1$
\begin{equation}
G(\mid\mathbf{r}-\mathbf{r}^{\prime}\mid,\beta,n\left(  \mathbf{r}\right)
)\rightarrow\frac{2\sqrt{\pi}}{\lambda^{3}}\frac{\lambda}{\mid\mathbf{r}%
-\mathbf{r}^{\prime}\mid}e^{-4\pi\left(  \mid\mathbf{r}-\mathbf{r}^{\prime
}\mid/\lambda\right)  ^{2}}\;. \label{3.22}%
\end{equation}
In this limit, $G(\mid\mathbf{r}-\mathbf{r}^{\prime}\mid,\beta,n\left(
\mathbf{r}\right)  )$ becomes independent of the density.

In the opposite limit, $T=0$ K, $G(\mid\mathbf{r}-\mathbf{r}^{\prime}%
\mid,\beta,n\left(  \mathbf{r}\right)  )$ becomes
\begin{equation}
G(\mid\mathbf{r}-\mathbf{r}^{\prime}\mid,\beta,n)=\frac{\sqrt{\pi}}{\ell
_{F}^{3}}\left(  \frac{\ell_{F}}{\left\vert \mathbf{r}-\mathbf{r}^{\prime
}\right\vert }\right)  ^{2}j_{1}\left(  4\pi\frac{\mid\mathbf{r}%
-\mathbf{r}^{\prime}\mid}{\ell_{F}}\right)  \;, \label{3.23}%
\end{equation}
where $\ell_{F}$ is the density-dependent Fermi length and $j_{1}\left(
x\right)  $ is a spherical Bessel function
\begin{equation}
\ell_{F}=\frac{2\pi}{k_{F}},\hspace{0.2in}k_{F}=2\pi\left(  \frac{3n}{8\pi
}\right)  ^{1/3},\hspace{0.2in}j_{1}\left(  x\right)  =\frac{1}{x^{2}}\left(
\sin x-x\cos x\right)  \;. \label{3.24}%
\end{equation}
Use has been made of the fact that $\mu^{(0)}(\beta,n)\rightarrow\hbar
^{2}k_{F}^{2}/2m$ at zero temperature.

In summary, the approximations introduced in this section lead to the TF form
for the density except with a renormalized KS potential. That potential is
``smoothed'' over a length scale that depends on the temperature or degree of
degeneracy. In this same approximation, zeroth order in $\Delta\mu_{KS}$, the
non-interacting free energy of (\ref{3.17}) becomes the TF result, except
evaluated at the improved density.

\section{Singular external potentials}

\label{sec5}To see how the revised Thomas-Fermi form (\ref{7.8}) removes the
problem of the local form (\ref{2.5}) for singular attractive potentials,
consider the external potential from $N_{i}$ positive ions with charges
$Z_{\alpha}$ and positions $\left\{  \mathbf{R}_{\alpha}\right\}  $,
\begin{equation}
v(\mathbf{r})=\sum_{\alpha=1}^{N_{i}}\frac{-Z_{\alpha}e^{2}}{\left\vert
\mathbf{r}-\mathbf{R}_{\alpha}\right\vert }\;. \label{4.1}%
\end{equation}
Obviously singular at the ionic sites, $\left\{  \mathbf{R}_{\alpha}\right\}
$, its contribution to the renormalized KS potential
%(\ref{3.19}) is
(\ref{7.9}) is
\begin{align}
\widetilde{v}(\mathbf{r,}\beta &  \mid n)=\int d\mathbf{r}^{\prime
}G(\left\vert \mathbf{r}-\mathbf{r}^{\prime}\right\vert \mathbf{,}%
\beta,n\left(  \mathbf{r}\right)  )\sum_{\alpha=1}^{N_{i}}\frac{Z_{\alpha
}e^{2}}{\left\vert \mathbf{r}^{\prime}-\mathbf{R}_{\alpha}\right\vert
}\nonumber\\
&  =-\sum_{\alpha=1}^{N_{i}}Z_{\alpha}e^{2}\int_{0}^{\infty}dr^{\prime
}r^{\prime2}G(r^{\prime},\beta,n\left(  \mathbf{r}\right)  )\int
d\Omega^{\prime}\frac{1}{\left\vert \mathbf{r}-\mathbf{R}_{\alpha}%
-\mathbf{r}^{\prime}\right\vert }\;. \label{4.2}%
\end{align}
The angular integral can be performed to get
\begin{equation}
\widetilde{v}(\mathbf{r,}\beta\mid n)=-\sum_{\alpha=1}^{N_{i}}\frac{Z_{\alpha
}e^{2}}{\left\vert \mathbf{r}-\mathbf{R}_{\alpha}\right\vert }S(\left\vert
\mathbf{r}-\mathbf{R}_{\alpha}\right\vert ,n\left(  \mathbf{r}\right)  )\;,
\label{4.3}%
\end{equation}
where \newline%
\begin{align}
S(\left\vert \mathbf{r}-\mathbf{R}_{\alpha}\right\vert ,\beta,n\left(
\mathbf{r}\right)  )  &  =4\pi\left(  \int_{0}^{\left\vert \mathbf{r}%
-\mathbf{R}_{\alpha}\right\vert }dr^{\prime}r^{\prime2}G(r^{\prime}%
,\beta,n\left(  \mathbf{r}\right)  )\right. \nonumber\\
&  +\left.  \left\vert \mathbf{r}-\mathbf{R}_{\alpha}\right\vert
\int_{\left\vert \mathbf{r}-\mathbf{R}_{\alpha}\right\vert }^{\infty
}dr^{\prime}r^{\prime}G(r^{\prime},\beta,n\left(  \mathbf{r}\right)  )\right)
\nonumber
\end{align}%
\begin{align}
&  =\frac{\left\vert \mathbf{r}-\mathbf{R}_{\alpha}\right\vert }{\lambda}%
\int_{0}^{\infty}dx\,p(x,\beta,\mu^{(0)}\left(  \beta,n\left(  \mathbf{r}%
\right)  \right)  )\left[  \left(  \frac{\lambda}{\left\vert \mathbf{r}%
-\mathbf{R}_{\alpha}\right\vert }\left(  1-\cos\left(  \frac{4x\sqrt{\pi
}\left\vert \mathbf{r}-\mathbf{R}_{\alpha}\right\vert }{\lambda}\right)
\right)  \right)  \right. \nonumber\\
&  \left.  +4x\sqrt{\pi}\left(  \frac{1}{2}\pi-\text{{Si}}\left(
\frac{4x\sqrt{\pi}\left\vert \mathbf{r}-\mathbf{R}_{\alpha}\right\vert
}{\lambda}\right)  \right)  \right]  \label{4.3a}%
\end{align}
and Si$\left(  x\right)  $ is the Sine integral
\begin{equation}
\text{Si}\left(  x\right)  =\int_{0}^{x}dx^{\prime}\frac{\sin x^{\prime}%
}{x^{\prime}}\;. \label{4.5}%
\end{equation}
Note also that in (\ref{4.3a}) $x$ is an integration variable, not a Cartesian
component of $\mathbf{r}$.

For $\left\vert \mathbf{r}-\mathbf{R}_{\alpha}\right\vert >>\lambda$,
$S(\left\vert \mathbf{r}-\mathbf{R}_{\alpha}\right\vert ,\beta,n)\rightarrow1$
and the Coulomb form for the potentials is recovered. However, for $\left\vert
\mathbf{r}-\mathbf{R}_{\alpha}\right\vert <<\lambda$, one has
%\begin{equation}
%S(\left\vert \mathbf{r}-\mathbf{R}_{\alpha}\right\vert ,\beta,n)\rightarrow
%2\pi^{3/2}\int_{0}^{\infty}dx\,xp(x,z)\frac{\left\vert \mathbf{r}%
%-\mathbf{R}_{\alpha}\right\vert }{\lambda}\;, \label{4.6}%
%\end{equation}%
\begin{equation}
S(\left\vert \mathbf{r}-\mathbf{R}_{\alpha}\right\vert ,\beta,n)\rightarrow
2\pi^{3/2}\frac{\left\vert \mathbf{r}-\mathbf{R}_{\alpha}\right\vert }%
{\lambda}\int_{0}^{\infty}dx\,xp(x,z)\;, \label{4.6}%
\end{equation}
and the Coulomb singularity is removed. At zero temperature, the length scale
$\lambda$ no longer is relevant and (\ref{4.3a}) becomes instead
\begin{align}
S(\left\vert \mathbf{r}-\mathbf{R}_{\alpha}\right\vert ,\beta,n)  &
\mid_{\beta=\infty}=1-\frac{\ell_{F}}{4\pi\left\vert \mathbf{r}-\mathbf{R}%
_{\alpha}\right\vert }\sin\left(  \frac{4\pi\left\vert \mathbf{r}%
-\mathbf{R}_{\alpha}\right\vert }{\ell_{F}}\right) \nonumber\\
&  +\frac{1}{2}\frac{4\pi\left\vert \mathbf{r}-\mathbf{R}_{\alpha}\right\vert
}{\ell_{F}}\left(  j_{1}\left(  \frac{4\pi\left\vert \mathbf{r}-\mathbf{R}%
_{\alpha}\right\vert }{\ell_{F}}\right)  +\frac{1}{2}\pi-\text{Si}\left(
\frac{4\pi\left\vert \mathbf{r}-\mathbf{R}_{\alpha}\right\vert }{\ell_{F}%
}\right)  \right)  \;. \label{4.7}%
\end{align}
One sees that $S(\left\vert \mathbf{r}-\mathbf{R}_{\alpha}\right\vert
,\beta,n\left(  \mathbf{r}\right)  )\mid_{\beta=\infty}\rightarrow1$ for
distances large compared to the Fermi length, and near $\left\vert
\mathbf{r}-\mathbf{R}_{\alpha}\right\vert =0$ behaves as
\begin{equation}
S(\left\vert \mathbf{r}-\mathbf{R}_{\alpha}\right\vert ,\beta,n)\mid
_{\beta=\infty}\rightarrow\frac{\left\vert \mathbf{r}-\mathbf{R}_{\alpha
}\right\vert }{\ell_{F}}\;. \label{4.8}%
\end{equation}
The singularity again is removed.

\section{Parr-Ghosh TF regularization}

\label{sec6}As noted in the Introduction, the problem of singular densities
within the TF approximation was addressed some time ago within zero
temperature DFT. The resolution given then is somewhat different from that
given here. Our analysis essentially extends the local TF approximation for
the density to include non-local effects necessary to smooth the singularity.
The earlier work of Parr and Ghosh \cite{PG86,GhoshParr87} addressed the
problem instead within the context of the standard TF functionals, but
constrained the class of densities to be considered. That type of analysis can
be extended to finite temperatures also, as illustrated in the following.

Return to the solution to the Euler equation in the TF approximation
(\ref{2.9})
\begin{equation}
\frac{\delta F_{TF}^{(0)}[n]}{\delta n(\mathbf{r})}=\mu(\mathbf{r}%
)-\frac{\delta F^{(1)}[n]}{\delta n(\mathbf{r})}. \label{5.1}%
\end{equation}
Consider again the case of an external potential due to $N_{i}$ ions,
(\ref{4.1}). Then, in addition to the constraints of $n(\mathbf{r})\geq0$ and
total number of electrons $\int d\mathbf{r}n(\mathbf{r})=N\left(  \mu
_{e}\right)  $, include the additional constraints
\cite{PG86,GoldsteinRieder87}
\begin{equation}
\int d\mathbf{r}\left\vert \mathbf{\nabla}_{\mathbf{r}-\mathbf{R}_{\alpha}%
}n(\mathbf{r})\right\vert ^{2}=\text{finite} \; . \label{5.2}%
\end{equation}
This can be accomplished by introducing a corresponding term in the free
energy, $\sum_{\alpha=1}^{N_{i}}\lambda_{\alpha}\int d\mathbf{r}\left\vert
\mathbf{\nabla}_{\mathbf{r}-\mathbf{R}_{\alpha}}n(\mathbf{r})\right\vert ^{2}%
$, where $\lambda_{\alpha}$ are Lagrange multipliers%
\begin{equation}
F[n]\rightarrow F[n]-\sum_{\alpha=1}^{N_{i}}\lambda_{\alpha}\int
d\mathbf{r}\left\vert \mathbf{\nabla}_{\mathbf{r}-\mathbf{R}_{\alpha}%
}n(\mathbf{r})\right\vert ^{2}. \label{5.3}%
\end{equation}
The Euler equation then becomes%

\begin{align}
\frac{\delta F_{TF}^{(0)}[n]}{\delta n(\mathbf{r})}  &  =\mu_{e}+\sum
_{\alpha=1}^{N_{i}}\left(  \frac{Z_{\alpha}e^{2}}{\left\vert \mathbf{r}%
-\mathbf{R}_{\alpha}\right\vert }+\lambda_{\alpha}\mathbf{\nabla}%
_{\mathbf{r}-\mathbf{R}_{\alpha}}^{2}n(\mathbf{r})\right)  -\frac{\delta
F^{(1)}[n]}{\delta n(\mathbf{r})}\nonumber\\
&  =\mu_{e}+\sum_{\alpha=1}^{N_{i}}\left(  \frac{Z_{\alpha}e^{2}%
-2\lambda_{\alpha}\mathbf{\partial}_{\left\vert \mathbf{r}-\mathbf{R}_{\alpha
}\right\vert }n(r)}{\left\vert \mathbf{r}-\mathbf{R}_{\alpha}\right\vert
}+\lambda_{\alpha}\mathbf{\partial}_{\left\vert \mathbf{r}-\mathbf{R}_{\alpha
}\right\vert }^{2}n(\mathbf{r})\right)  -\frac{\delta F^{(1)}[n]}{\delta
n(\mathbf{r})} \label{5.4}%
\end{align}
In the vicinity of $\mathbf{R}_{\alpha}$, assume that $n(\mathbf{r})$ depends
only on the relative radial coordinate, i.e. $n(\mathbf{r})\rightarrow
\overline{n}_{\alpha}(\left\vert \mathbf{r}-\mathbf{R}_{\alpha}\right\vert
)$.\texttt{ }Then, as $\mathbf{r}\rightarrow\mathbf{R}_{\alpha}$%
\begin{equation}
\frac{Z_{\alpha}e^{2}}{\left\vert \mathbf{r}-\mathbf{R}_{\alpha}\right\vert
}+\lambda_{\alpha}\mathbf{\nabla}_{\mathbf{r}-\mathbf{R}_{\alpha}}%
^{2}n(\mathbf{r})\rightarrow\frac{Z_{\alpha}e^{2}-2\lambda_{\alpha}%
\overline{n}_{\alpha}^{\prime}}{\left\vert \mathbf{r}-\mathbf{R}_{\alpha
}\right\vert }+\lambda_{\alpha}\overline{n}_{\alpha}^{\prime\prime}\mathbf{,}
\label{5.5}%
\end{equation}
where $\overline{n}_{\alpha}^{\prime}$ and $\overline{n}_{\alpha}%
^{\prime\prime}$ are the first and second derivatives of $\overline{n}%
_{\alpha}(x)$. The singularity is therefore removed by the choice%
\begin{equation}
\lambda_{\alpha}=\frac{Z_{\alpha}e^{2}}{2\overline{n}_{\alpha}^{\prime}}.
\label{5.6}%
\end{equation}
This is only implicit since $\overline{n}_{\alpha}^{\prime}$ is not known
\textit{a priori}. Hence the non-singular Euler equation,
\begin{equation}
\frac{\delta F_{TF}^{(0)}[n]}{\delta n(\mathbf{r})}=\mu_{e}+\sum_{\alpha
=1}^{N_{i}}\left(  \frac{Z_{\alpha}e^{2}-2\lambda_{\alpha}\mathbf{\partial
}_{\left\vert \mathbf{r}-\mathbf{R}_{\alpha}\right\vert }n(r)}{\left\vert
\mathbf{r}-\mathbf{R}_{\alpha}\right\vert }+\lambda_{\alpha}\mathbf{\partial
}_{\left\vert \mathbf{r}-\mathbf{R}_{\alpha}\right\vert }^{2}n(r)\right)
-\frac{\delta F^{(1)}[n]}{\delta n(\mathbf{r})}, \label{5.7}%
\end{equation}
must be solved self-consistently with (\ref{5.6}).

\section{Discussion}

\label{sec7}

Consideration of the relationship of our approach to that of Parr and Ghosh
illustrates the difference between revising the TF scheme as we have done and
amending (or repairing) it. The comparison proceeds as follows. The Euler
equation of Section \ref{sec3} in the form (\ref{3.5}) is
\begin{equation}
\mu^{(0)}(\beta,n\left(  \mathbf{r}\right)  )=\mu_{e}-v_{KS}(\mathbf{r,}%
\beta\mid n)-\int d\mathbf{r}^{\prime}G(\left\vert \mathbf{r}-\mathbf{r}%
^{\prime}\right\vert \mathbf{,}\beta,n\left(  \mathbf{r}\right)  )\left[
v_{KS}(\mathbf{r}^{\prime}\mathbf{,}\beta\mid n)-v_{KS}(\mathbf{r,}\beta\mid
n)\right]  , \label{5.8}%
\end{equation}
while the corresponding Parr-Ghosh type equation (\ref{5.7}) is
\begin{equation}
\mu^{(0)}(\beta,n\left(  \mathbf{r}\right)  )=\mu_{e}-v_{KS}(\mathbf{r,}%
\beta\mid n)-\sum_{\alpha=1}^{N_{i}}\lambda_{\alpha}\mathbf{\nabla
}_{\mathbf{r}-\mathbf{R}_{\alpha}}^{2}n(\mathbf{r}). \label{5.9}%
\end{equation}
Both provide non-local corrections to the local TF Euler equation to remove
the singularity. The Parr-Ghosh form involves terms through second order in
the gradients while the approach here is fully non-local. Both require changes
in the TF free energy functional. The Parr-Ghosh form is an additive
contribution from the Lagrange multiplier, while here the modification is via
global reweighting through $G(\mathbf{r},\mathbf{r}^{\prime}\mathbf{,}%
\beta\mid u) $.

For at least two reasons, the present approach seems more systematic and
general than previous ones. First, it does not involve \textit{ad hoc} choices
of constraints or imposition of repairs. Second, it is valid at all
temperatures. As noted in the introduction, the approach here is a
formalization and extension of the idea proposed by Harris, Hoffman, and Pratt
\ \cite{PrattHoffmanHarris,HoffmanHarrisPratt}. In fact, eqs.\ (10) and (11)
of Ref. \cite{HoffmanHarrisPratt} are effectively the same as the simple
approximation of section \ref{sec4} here for the density.

In retrospect it is interesting to reflect on the TF limit itself. That
corresponds to the choice $z(\mathbf{r,}\beta\mid\mu_{KS})=\mu_{KS}%
(\mathbf{r})$. However, the exact result (\ref{3.10}) gives for this choice
\begin{equation}
\int d\mathbf{r}^{\prime}G(\mathbf{r},\mathbf{r}^{\prime}\mathbf{,}\beta\mid
z,\Delta\mu_{KS})\left(  \mu_{KS}(\mathbf{r}^{\prime})-\mu_{KS}(\mathbf{r}%
)\right)  =0, \label{5.10}%
\end{equation}
which implies that the system must be uniform over length scales of the
response function.

\section{Acknowledgment}

This research was supported by US DOE Grant DE-SC0002139.

\bigskip

\appendix

\section{Non-interacting functionals}

\label{apa}The starting point for the non-interacting functionals in
statistical mechanics is the grand potential $\Omega^{(0)}(\beta\mid\mu)$ for
the grand ensemble. Its evaluation leads to the single particle form
\cite{DuftyTrickey11}%
\begin{equation}
\Omega^{(0)}(\beta\mid\mu)=-\int d\mathbf{r}\beta^{-1}\left\langle
\mathbf{r}\right\vert \ln\left(  1+e^{-\beta\left(  \frac{\widehat{p}^{2}}%
{2m}-\mu(\widehat{\mathbf{r}})\right)  }\right)  \left\vert \mathbf{r}%
\right\rangle \label{B.1}%
\end{equation}
where $\left\langle \mathbf{r}\right\vert \widehat{X}\left\vert \mathbf{r}%
\right\rangle $ denotes the diagonal matrix element of the operator
$\widehat{X}$ in coordinate representation. The local chemical potential is
given by $\mu(\mathbf{r})\equiv\mu_{e}-v(\mathbf{r})$, where $v(\mathbf{r})$
is the given external potential. The conjugate thermodynamic variable is the
density, defined by%
\begin{equation}
\frac{\delta\Omega^{(0)}(\beta\mid\mu)}{\delta\mu\left(  \mathbf{r}\right)
}=-n^{(0)}\left(  \mathbf{r}\right)  =-\left\langle \mathbf{r}\right\vert
\left(  e^{\beta\left(  \frac{\widehat{p}^{2}}{2m}-\mu(\widehat{\mathbf{r}%
})\right)  }+1\right)  ^{-1}\left\vert \mathbf{r}\right\rangle , \label{B.2}%
\end{equation}
which follows by direct calculation. The non-interacting free energy is
defined by the Legendre transform%
\[
F^{(0)}(\beta\mid n^{(0)})\equiv\Omega^{(0)}(\beta\mid\mu)+\int d\mathbf{r}%
\mu(\mathbf{r})n^{(0)}\left(  \mathbf{r}\right)
\]%
\begin{equation}
=\int d\mathbf{r}\left[  -\beta^{-1}\left\langle \mathbf{r}\right\vert
\ln\left(  1+e^{-\beta\left(  \frac{\widehat{p}^{2}}{2m}-\mu(\widehat
{\mathbf{r}})\right)  }\right)  \left\vert \mathbf{r}\right\rangle
+\mu(\mathbf{r})n^{(0)}(\mathbf{r})\right]  . \label{B.3}%
\end{equation}
This definition gives the relation%
\begin{equation}
\frac{\delta F^{(0)}(\beta\mid n^{(0)})}{\delta n^{(0)}\left(  \mathbf{r}%
\right)  }=\mu\left(  \mathbf{r}\right)  . \label{B.5}%
\end{equation}

These equations give the non-interacting free energy as a functional of
$n^{(0)}\left(  \mathbf{r}\right)  $, the density following from
$\mu(\mathbf{r})$ for the non-interacting system. However, in the main text
that free energy functional is required at the density $n\left(
\mathbf{r}\right)  $ for the interacting system. That requirement is met by
replacing $\mu(\mathbf{r})$ by the Kohn-Sham local chemical potential
$\mu_{KS}(\mathbf{r})$ given in Eq.\ (\ref{2.3a}) in these non-interacting
functional expressions.

\section{Singularity of density}

\label{apb} The dimensionless form of Eq.\ (\ref{2.5}) written in terms of
standard Fermi-Dirac integrals is
\begin{equation}
\lambda^{3}n_{TF}(\mathbf{r})=2f_{3/2}(e^{\beta\mu_{KS}(\mathbf{r},\beta\mid
n)})\;, \label{dimensionlessden}%
\end{equation}
with $\lambda=h\sqrt{\beta/2\pi m}$ the thermal de Broglie wavelength. For
large $x$
\begin{equation}
f_{3/2}(x)\rightarrow\frac{4}{3\sqrt{\pi}}(\ln(x))^{3/2}\;.
\label{limf3halves}%
\end{equation}
Sufficiently close to the negative singularity of the attractive Coulomb
potential, $e^{\beta(\mu_{KS}(\mathbf{r},\beta\mid n)-\mu_{e})}\gg1$, so
(\ref{dimensionlessden}) goes as
\begin{equation}
n_{TF}(\mathbf{r})\rightarrow\frac{8}{3\sqrt{\pi}}\left(  \frac{m}{2\pi
\hbar^{2}}\right)  ^{3/2}\mu_{KS}^{3/2}(\mathbf{r},\beta\mid n)=\left(
\frac{3}{5c_{TF}}\right)  ^{3/2}\left(  \mu_{e}-v_{KS}(\mathbf{r,}\beta\mid
n)\right)  ^{3/2}\;.\nonumber
\end{equation}
This agrees with Eq.\ (\ref{nTFatTeq0}).

\section{Formal representations for $z(\mathbf{r})$ and $F^{(0)}$}

\label{apc} In this Appendix $z(\mathbf{r,}\beta\mid\mu_{KS})$ and
$F^{(0)}\left(  \beta\mid n\right)  $ are written in terms of the non-linear
response to the spatial variations of $\mu_{KS}$\textbf{ }relative to $z$ at a
particular point, $\Delta\mu_{KS}\left(  \mathbf{r}^{\prime}\right)  $. Recall
from (\ref{3.3}) that this is a measure of the non-uniformity of the system.
Consider first the density and write it is%

\begin{equation}
n(\mathbf{r},\beta\mid\mu_{KS})=\left\langle \mathbf{r}\right\vert \left(
e^{\left(  \beta\left(  \frac{\widehat{p}^{2}}{2m}-\mu_{KS}(\mathbf{\widehat
{\mathbf{q}},}\beta\mid n)\right)  \right)  }+1\right)  ^{-1}\left\vert
\mathbf{r}\right\rangle =\left\langle \mathbf{r}\right\vert \left(  e^{\left(
\beta\left(  \frac{\widehat{p}^{2}}{2m}-z(\mathbf{r})-\Delta\mu_{KS}%
(\mathbf{\widehat{\mathbf{q}},}\beta\mid n)\right)  \right)  }+1\right)
^{-1}\left\vert \mathbf{r}\right\rangle \label{c.0}%
\end{equation}
with%
\begin{equation}
\Delta\mu_{KS}\left(  \mathbf{\widehat{\mathbf{q}},}\beta\mid n\right)
=\mu_{KS}\left(  \mathbf{\widehat{\mathbf{q}}}\right)  -z(\mathbf{r,}\beta
\mid\mu_{KS}). \label{c.1}%
\end{equation}
As shown in (\ref{7.1}), $\Delta\mu_{KS}\left(  \mathbf{q,}\beta\mid n\right)
$ is a tempered measure of the non-uniformity of $\mu_{KS}\left(
\mathbf{q,}\beta\mid n\right)  $. A formal representation in terms of
$\Delta\mu_{KS}$ is obtained from the identity
\begin{align}
n(\mathbf{r},\beta &  \mid\mu_{KS})=n(\mathbf{r},\beta\mid z)+\int_{0}%
^{1}dx\frac{d}{dx}n(\mathbf{r},\beta\mid z+x\Delta\mu_{KS})\nonumber\\
&  =n(\mathbf{r},\beta\mid z)+\int_{0}^{1}dx\int d\mathbf{r}^{\prime}%
\frac{\delta n(\mathbf{r},\beta\mid u)}{\delta u\left(  \mathbf{r}^{\prime
}\right)  }\mid_{z+x\Delta\mu_{KS}}\left(  \mu_{KS}\left(  \mathbf{r}^{\prime
}\right)  -z\left(  \mathbf{r}\right)  \right)  \label{c.2}%
\end{align}
Since $z$ is the \emph{fixed} value $z\left(  \mathbf{r}\right)  $ at all
points $\mathbf{r}^{\prime}$, the first term of (\ref{c.2}) becomes the TF
form (\ref{3.3}) and is equal to $n(\mathbf{r})$. As a consequence, the second
term of (\ref{c.2}) must vanish
\begin{equation}
\int_{0}^{1}dx\int d\mathbf{r}^{\prime}\frac{\delta n(\mathbf{r},\beta\mid
u)}{\delta u\left(  \mathbf{r}^{\prime}\right)  }\mid_{z+x\Delta\mu_{KS}%
}\left(  \mu_{KS}\left(  \mathbf{r}^{\prime}\right)  -z\left(  \mathbf{r}%
\right)  \right)  =0\;. \label{c.3}%
\end{equation}
Equation\ (\ref{c.3}) gives the formally exact representation for $z\left(
\mathbf{r}\right)  $ as a functional of $\mu_{KS}$
\begin{equation}
z(\mathbf{r,}\beta\mid\mu_{KS})=\int d\mathbf{r}^{\prime}G(\mathbf{r}%
,\mathbf{r}^{\prime}\mathbf{,}\beta\mid z,\mu_{KS})\mu_{KS}(\mathbf{r}%
^{\prime})\;. \label{c.4}%
\end{equation}
Here $G(\mathbf{r},\mathbf{r}^{\prime}\mathbf{,}\beta\mid z,\mu_{KS})$ is
constructed from the ideal Fermi gas response functional in terms of
$\Delta\mu_{KS}$ as described by (\ref{3.11}) and (\ref{3.12}).

Next the grand potential $\Omega^{(0)}(\beta\mid\mu_{KS})$ is written in terms
of its corresponding density $\omega^{(0)}(\mathbf{r},\beta\mid\mu)$%
\begin{equation}
\Omega^{(0)}(\beta\mid\mu)\equiv\int d\mathbf{r}\omega^{(0)}(\mathbf{r}%
,\beta\mid\mu). \label{c.5}%
\end{equation}
Then a corresponding representation in terms of $\Delta\mu_{KS}$ follows from
the identity corresponding to (\ref{c.2})%

\begin{align}
\omega^{(0)}(\mathbf{r},\beta &  \mid\mu_{KS})=\omega^{(0)}(\mathbf{r}%
,\beta\mid z)+\int_{0}^{1}dx\frac{d}{dx}\omega^{(0)}(\mathbf{r},\beta\mid
z+x\Delta\mu_{KS})\nonumber\\
&  =\omega_{TF}(\beta,z(\mathbf{r,}\beta\mid\mu_{KS}))+\int_{0}^{1}dx\int
d\mathbf{r}^{\prime}\frac{\delta\omega^{(0)}(\mathbf{r},\beta\mid u)}{\delta
u\left(  \mathbf{r}^{\prime}\right)  }\mid_{z+x\Delta\mu_{KS}}\Delta\mu
_{KS}\left(  \mathbf{r}^{\prime}\right)  . \label{c.6}%
\end{align}
The first term on the right is the TF result (local density approximation)
since%
\begin{align}
\omega^{(0)}(\mathbf{r},\beta &  \mid z)=-\beta^{-1}\left\langle
\mathbf{r}\right\vert \ln\left(  1+e^{-\beta\left(  \frac{\widehat{p}^{2}}%
{2m}-z(\mathbf{r,}\beta\mid\mu_{KS})\right)  }\right)  \left\vert
\mathbf{r}\right\rangle \nonumber\\
&  =-\frac{2}{\beta h^{3}}\int d\mathbf{p}\ln\left(  1+e^{-\beta\left(
\frac{p^{2}}{2m}-z(\mathbf{r,}\beta\mid\mu_{KS})\right)  }\right)
\equiv\omega_{TF}^{(0)}(\beta,z(\mathbf{r,}\beta\mid\mu_{KS})). \label{c.7}%
\end{align}
Rewrite the second term of (\ref{c.6}) with the identity%
\[
\int_{0}^{1}dx\frac{\delta\omega^{(0)}(\mathbf{r},\beta\mid u)}{\delta
u\left(  \mathbf{r}^{\prime}\right)  }\mid_{x\Delta\mu_{KS}+z}=\frac
{\delta\omega^{(0)}(\mathbf{r},\beta\mid u)}{\delta u\left(  \mathbf{r}%
^{\prime}\right)  }\mid_{z\left(  \mathbf{r}\right)  }+\int_{0}^{1}dx\int
_{0}^{x}dx^{\prime}\frac{d}{dx^{\prime}}\frac{\delta\omega^{(0)}%
(\mathbf{r},\beta\mid u)}{\delta u\left(  \mathbf{r}^{\prime}\right)  }%
\mid_{x^{\prime}\Delta\mu_{KS}+z}%
\]%
\begin{align}
&  =\frac{\delta\omega^{(0)}(\mathbf{r},\beta\mid u)}{\delta u\left(
\mathbf{r}^{\prime}\right)  }\mid_{z}+\int_{0}^{1}dx\int_{0}^{x}dx^{\prime
}\int d\mathbf{r}^{\prime\prime}\frac{\delta^{2}\omega^{(0)}(\mathbf{r}%
,\beta\mid u)}{\delta u\left(  \mathbf{r}^{\prime}\right)  \delta u\left(
\mathbf{r}^{\prime\prime}\right)  }\mid_{x^{\prime}\Delta\mu_{KS}+z}\Delta
\mu_{KS}\left(  \mathbf{r}^{\prime\prime}\right) \nonumber\\
&  =\frac{\delta\omega^{(0)}(\mathbf{r},\beta\mid u)}{\delta u\left(
\mathbf{r}^{\prime}\right)  }\mid_{z}+\int_{0}^{1}dx\left(  1-x\right)  \int
d\mathbf{r}^{\prime\prime}\frac{\delta^{2}\omega^{(0)}(\mathbf{r},\beta\mid
u)}{\delta u\left(  \mathbf{r}^{\prime}\right)  \delta u\left(  \mathbf{r}%
^{\prime\prime}\right)  }\mid_{x\Delta\mu_{KS}+z}\Delta\mu_{KS}\left(
\mathbf{r}^{\prime\prime}\right)  . \label{c.8}%
\end{align}
Also, note that%
\begin{equation}
\int d\mathbf{r}^{\prime}\frac{\delta\omega^{(0)}(\mathbf{r},\beta\mid
u)}{\delta u\left(  \mathbf{r}^{\prime}\right)  }\mid_{z}=\frac{\partial
\omega_{TF}(\beta,z(\mathbf{r,}\beta\mid\mu_{KS}))}{\partial z(\mathbf{r,}%
\beta\mid\mu_{KS})}=-n_{TF}(\beta,z(\mathbf{r,}\beta\mid\mu_{KS})).
\label{c.9}%
\end{equation}
With these results (\ref{c.7}) becomes%
\[
\omega^{(0)}(\mathbf{r},\beta\mid\mu_{KS})=\omega_{TF}(\beta,z(\mathbf{r,}%
\beta\mid\mu_{KS}))\ +n_{TF}(\beta,z(\mathbf{r,}\beta\mid\mu_{KS}%
))z(\mathbf{r,}\beta\mid\mu_{KS})
\]%
\begin{equation}
+\int d\mathbf{r}^{\prime}\frac{\delta\omega^{(0)}(\mathbf{r},\beta\mid
z)}{\delta z\left(  \mathbf{r}^{\prime}\right)  }\mu_{KS}\left(
\mathbf{r}^{\prime}\right)  +\int d\mathbf{r}^{\prime}d\mathbf{r}%
^{\prime\prime}h(\mathbf{r};\mathbf{r}^{\prime},\mathbf{r}^{\prime\prime
},\beta\mid\Delta\mu_{KS})\Delta\mu_{KS}\left(  \mathbf{r}^{\prime}\right)
\Delta\mu_{KS}\left(  \mathbf{r}^{\prime\prime}\right)  \label{c.10}%
\end{equation}
where the non-linear response function $h(\mathbf{r};\mathbf{r}^{\prime
},\mathbf{r}^{\prime\prime},\beta\mid z,\Delta\mu_{KS})$ is
\begin{equation}
h(\mathbf{r};\mathbf{r}^{\prime},\mathbf{r}^{\prime\prime},\beta\mid
z,\Delta\mu_{KS})\equiv\int_{0}^{1}dx\left(  1-x\right)  \frac{\delta
^{2}\omega^{(0)}(\mathbf{r},\beta\mid u)}{\delta u\left(  \mathbf{r}^{\prime
}\right)  \delta u\left(  \mathbf{r}^{\prime\prime}\right)  }\mid_{x\Delta
\mu_{KS}+z}. \label{c.11}%
\end{equation}
Finally, integrating over $\mathbf{r}$ gives the non-interacting grand
potential in the form%
\[
\Omega^{(0)}(\beta\mid\mu_{KS})=\Omega_{TF}(\beta\mid z)-\int d\mathbf{r}%
n_{TF}(\beta,z(\mathbf{r,}\beta\mid\mu_{KS}))\Delta\mu_{KS}\left(
\mathbf{r}\right)
\]%
\begin{equation}
+\int d\mathbf{r}d\mathbf{r}^{\prime}d\mathbf{r}^{\prime\prime}h(\mathbf{r}%
;\mathbf{r}^{\prime},\mathbf{r}^{\prime\prime},\beta\mid z,\Delta\mu
_{KS})\Delta\mu_{KS}\left(  \mathbf{r}^{\prime}\right)  \Delta\mu_{KS}\left(
\mathbf{r}^{\prime\prime}\right)  . \label{c.12}%
\end{equation}

The non-interacting free energy follows directly from its definition as the
Legendre transform%
\begin{align}
F^{(0)}(\beta &  \mid n)\equiv\Omega^{(0)}(\beta\mid\mu_{KS})+\int
d\mathbf{r}n_{TF}(\beta,z(\mathbf{r,}\beta\mid\mu_{KS}))\mu_{KS}\left(
\mathbf{r}\right) \nonumber\\
&  =F_{TF}(\beta\mid n)+\int d\mathbf{r}d\mathbf{r}^{\prime}d\mathbf{r}%
^{\prime\prime}h(\mathbf{r};\mathbf{r}^{\prime},\mathbf{r}^{\prime\prime
},\beta\mid z,\Delta\mu_{KS})\Delta\mu_{KS}\left(  \mathbf{r}^{\prime}\right)
\Delta\mu_{KS}\left(  \mathbf{r}^{\prime\prime}\right)  . \label{c.13}%
\end{align}
Here%
\begin{equation}
F_{TF}(\beta\mid n)=\Omega_{TF}(\beta\mid z)+\int d\mathbf{r}n_{TF}%
(\beta,z(\mathbf{r}))z\left(  \mathbf{r}\right)  , \label{c.13a}%
\end{equation}
This is the result quoted in the text.

\section{Response functions $R(r\mathbf{,}\beta,n)$ and $G(r\mathbf{,}%
\beta,n)$}

\label{apd} The normalized response function $R(r\mathbf{,}\beta,n)$ is
defined in terms of the ideal Fermi gas response function
\begin{equation}
R(\left\vert \mathbf{r}-\mathbf{r}^{\prime}\right\vert \mathbf{,}%
\beta,z)=\frac{\delta n(\mathbf{r},\beta\mid x)}{\delta x\left(
\mathbf{r}^{\prime}\right)  }\mid_{x=z}=-\int_{0}^{\beta}dy\left\langle
e^{y\left(  \widehat{H}_{N}-zN\right)  }\widehat{n}(\mathbf{r}^{\prime
})e^{-y\left(  \widehat{H}_{N}-zN\right)  }\widehat{n}(\mathbf{r}%
)\right\rangle -n^{2}. \label{a.1}%
\end{equation}
The Hamiltonian operator $\widehat{H}_{N}$ is that for a uniform ideal Fermi
gas, $\widehat{n}(\mathbf{r})$ is the number density operator, and the
brackets $\left\langle ..\right\rangle $ denote an average over the associated
grand canonical ensemble. The calculation is straightforward leading to%
\begin{equation}
R(r\mathbf{,}\beta,z)=-\left(  2s+1\right)  \int\frac{d\mathbf{k}}{\left(
2\pi\right)  ^{3}}e^{-i\mathbf{k\cdot r}}\int\frac{d\mathbf{k}_{1}}{\left(
2\pi\right)  ^{3}}\frac{\left(  n\left(  \epsilon_{\left\vert \mathbf{k-k}%
_{1}\right\vert }\right)  -n\left(  \epsilon_{k_{1}}\right)  \right)
}{\epsilon_{k_{1}}-\epsilon_{\left\vert \mathbf{k-k}_{1}\right\vert }},
\label{a.2}%
\end{equation}
where $s$ is the particle spin, $\epsilon_{k}=\hbar^{2}k^{2}/2m$, and
\begin{equation}
n\left(  \epsilon_{k}\right)  =\left(  e^{\beta\left(  \epsilon_{k}-z\right)
}+1\right)  ^{-1}. \label{a.3}%
\end{equation}
Use cylindrical coordinates with the $z$ axis along $\mathbf{k}$ to reduce
(\ref{a.2}) further to%
\begin{equation}
R(r\mathbf{,}\beta,z)=-\left(  2s+1\right)  \frac{\beta}{k\lambda^{4}}%
\int\frac{d\mathbf{k}}{\left(  2\pi\right)  ^{3}}e^{-i\mathbf{k\cdot r}}%
\int_{-\infty}^{\infty}dx\frac{1}{x-\kappa/2}\ln\left(  \frac{1+e^{\beta
z}e^{-x^{2}}}{1+e^{\beta z}e^{-\left(  x-\kappa\right)  ^{2}}}\right)
\label{a.4}%
\end{equation}
where $\mathbf{\kappa=k}\lambda/2\sqrt{\pi}$ and $\lambda=\left(  2\pi
\beta\hbar^{2}/m\right)  ^{1/2}$. Finally,
\begin{equation}
R(r\mathbf{,}\beta,z)=\left(  2s+1\right)  \frac{\beta}{\pi\lambda^{4}}%
\frac{1}{r^{2}}\int_{0}^{\infty}dxx\left(  e^{-\beta z}e^{x^{2}}+1\right)
^{-1}\sin\left(  x\frac{4\sqrt{\pi}r}{\lambda}\right)  . \label{a.5}%
\end{equation}

To construct $G(r\mathbf{,}\beta,n)$ from (\ref{3.15}) the integral of
$R(r\mathbf{,}\beta,z)$ is needed%
\begin{equation}
\int d\mathbf{r}R(r\mathbf{,}\beta,z)=4\pi\left(  2s+1\right)  \frac{\beta
}{\pi\lambda^{4}}\int_{0}^{\infty}dx\left(  e^{-\beta z}e^{x^{2}}+1\right)
^{-1}\frac{\lambda}{4\sqrt{\pi}}. \label{a.6}%
\end{equation}
Then%
\begin{equation}
G(r\mathbf{,}\beta,n)=\frac{1}{\lambda\sqrt{\pi}}\frac{1}{r^{2}}\int
_{0}^{\infty}dxxp(x,\beta,z)\sin\left(  x\frac{4\sqrt{\pi}r}{\lambda}\right)
. \label{a.7}%
\end{equation}
with the normalized distribution%
\begin{equation}
p(x,\beta,z)\equiv\frac{\left(  e^{-\beta z}e^{x^{2}}+1\right)  ^{-1}}%
{\int_{0}^{\infty}dx\left(  e^{-\beta z}e^{x^{2}}+1\right)  ^{-1}} \label{a.8}%
\end{equation}

\bigskip
\end{document}